\documentclass[conference]{IEEEtran}
\IEEEoverridecommandlockouts
\usepackage[left=1.62cm,right=1.62cm,top=1.9cm]{geometry}
\usepackage{amsmath,amssymb,amsfonts}
\usepackage{algorithmic}
\usepackage{graphicx}
\usepackage{textcomp}
\usepackage{xcolor}
\usepackage{gnuplottex}
\usepackage{float}
\usepackage{subcaption}
\usepackage{svg}
\usepackage{pgf}
\usepackage{siunitx}
\usepackage{balance}
\usepackage[shortcuts,acronyms]{glossaries}
\usepackage[backend=biber, sorting=none]{biblatex}
\newacronym{RCU}{RCU}{Robot-Compute Unit}
\newacronym{JDR}{JDR}{Joint Detection Receiver}
\newacronym{SSR}{SSR}{Single Shot Receiver}
\newacronym{OJDR}{OJDR}{Optimal Joint Detection Receiver}
\newacronym{OSSR}{OSSR}{Optimal Single Shot Receiver}
\newacronym{OEAR}{OEAR}{Optimal Entanglement-Assisted Receiver}
\newacronym{M2M}{M2M}{Machine to Machine}
\newacronym{M2C}{M2C}{Machine to Cloud}
\newacronym{CDMA}{CDMA}{Code Division Multiple Access}
\newacronym{QIP}{QIP}{Quantum Information Processing}
\newacronym{OFDM}{OFDM}{Orthogonal Frequency Division Multiple Access}

\title{Scaling of Entanglement-Assisted Communication in Multi-Mode Amplified Fiber Links}
\author{
    \IEEEauthorblockN{Sekavčnik, Simon and Nötzel, Janis}\\
    \IEEEauthorblockA{Theoretical Quantum Systems Design, Technical University of Munich \\\{simon.sekavcnik, janis.noetzel\}@tum.de}
}

\date{April 2023}

\begin{document}
\maketitle

\begin{abstract}
Quantum information processing technology offers several communication strategies, which offer capacity advantages over classical technologies. However, advantages typically arise only in very particular communication scenarios which are of limited use in public networks. Most importantly, striking capacity advantages have so far been found only for cases where the system capacity is way below commercially interesting values. In this work we present a novel scenario where pre-shared entanglement offers arbitrarily high capacity advantages, and where at the same time data rates are compatible with future network demand. Our approach rests on the observation that the number of modes in multi-mode fiber can be increased solely by tuning of the refractive index, while maintaining the fiber diameter.
\end{abstract}



\section{Introduction}

\IEEEPARstart{C}{ommunication} networks are becoming more complex, and our dependence on them is increasing rapidly. The total amount of network traffic is poised to increase. With 6G networks, we are promised various applications which can only be made available using ultra-low latency communications. The challenge we face is meeting outlined future demands with the constraint of transmission power limit imposed by the optical fiber links.

The vast majority of network traffic today is serviced by optical fiber links. With current classical technologies, the communication capacity and effective rates are limited by the Shannon capacity. The eventual emergence of \ac{QIP} technologies will, in addition to providing new, never seen before applications and communication protocols - move the theoretical communication capacity bounds.

The fact that capacity bounds for \ac{QIP} enabled communications are higher than classical communication techniques is well documented \cite{guha2020infinite, giovannetti2003broadband}. However, finding use cases where the use of \ac{QIP} would translate into significant and practical communication advantage is difficult. Especially difficult is fitting an entanglement-assisted communication scheme into a real-world communication problem. In \cite{ntzel2022operating, jarzyna2019ultimate}, the authors examined the use of \ac{QIP} for the transceiver design of optical fiber networks transceiver. They pointed out the possibility of reducing the energy consumption of optical amplifiers. In \cite{ntzel2022operating}, different capacity limits of amplified optical fiber limits were calculated, which accounted for different relations between transmission power and amplifier gains. The work \cite{amiri2022preprocessing} studied the performance under more realistic conditions, taking into account Kerr-induced phase noise based on the model \cite{kunz2018phasenoise} and a specific design of the transceiver based on the work \cite{guha11}. Later \cite{sekavcnik2022effects} tried to showcase the advantages of entanglement-assisted \ac{QIP} in high-density, ultra-low latency networks. The advantages were laid out; however, it was also pointed out that the parameter region required for meaningful entanglement-assisted \ac{QIP} advantages is challenging and hard to argue.

In this work, we focus on outlining a practical communication scenario in which the use of \ac{QIP}, particularly with \textit{entanglement assistance}, translates into a capacity advantage over \textit{purely classical} and \textit{non-assisted} \ac{QIP} method. In order to showcase the different capacity bounds, we consider an optical fiber link with three different transmitter and receiver types.

In our scenario, the communication channel models an optic fiber with attenuation $\tau$, which has the ability to transmit $M$ number of orthogonal modes. The transceiver transmits a signal with some power $P$ in the number of photons per channel used. The noise experienced by the receiver is denoted by $\nu$ also in the number of thermal noise photons per channel use. All three capacities (Shannon, Holevo, and entanglement-assisted capacity) can be calculated for such channels. The differences between the three capacities stem from the fact that when the number of photons at the receiver ($\tau P/M$) is small enough (for example, below one), the \ac{QIP} enabled system will display an advantage over the purely classical system. The advantage scales logarithmically with $M/\tau P$. Additionally, the entanglement assisted links will display additional advantage over non-assisted \ac{QIP} enabled links if there is a sufficient amount of noise in the system ($\nu \gg 1$), combined with transmitters operating with low power per pulse $P/M\ll 1$. This advantage again scales logarithmically with $M/P$ \cite{guha2020infinite}.

Our task is to understand the technological meaning and explore their potential impacts on future communication system design. Therefore we root our modeling in the current fiber transmission reality by assuming communication at \SI{1550}{\nano\meter}. Here the fiber nonlinearities dictate a power limit of approximately \SI{1}{\milli\watt} per channel, which at a given wavelength of \SI{1550}{\nano\meter} results at around \SI{1e16}{\per\second} photons emitted at the sender. Values for $B$ in the order of \SI{10}{\giga\hertz} to \SI{1}{\tera\hertz} can be achieved with current technology, leaving us with at least \num{1e3} photons per pulse. A regime where power per pulse is much less than the number of modes ($P/M \ll 1$) is thus difficult to justify with single mode fibers. The recent literature, however, points out the possibility of using multimode fiber to transmit several spatially orthogonal modes over the same fiber. A number $N = \num{1e3}$ of modes has already been demonstrated as technically feasible today \cite{essiambre2013mmf}, implying that the region $P/M\approx1$ can be reached. To justify even lower ratios of $P/M$ we note that the number of spatial modes $N$ per fiber is governed by the formula $N\approx 0.5 (2\pi \tfrac{r}{\lambda} \sqrt{n^2_\text{core}-n^2_\text{clad}})^2$ \cite{yoon2012experimental}, where $r$ is the radius of the fiber, $\lambda$ the carrier wavelength, and $n_\text{core}$ and $n_\text{clad}$ the refractive index of the core and cladding. 
Taking into account that both the energy per photon and the slot rate scale as $\lambda^{-1}$, the number of photons per mode scales as $\lambda^{-4}$. This suggests that transmission at lower wavelengths might decrease the ratio $P/M$ further.  

Considering such possibilities, we may set $M=N \cdot B$ to argue for the very low power per pulse needed to see the advantages of entanglement-assisted communication. Since optical amplification is the standard technology enabling high throughput data transmission over more than \SI{100}{\kilo\meter}, and since optical amplification not only regenerates the signal but inevitably also introduces additional noise to the signal, it is, therefore, possible to motivate communication in a setting where both $\nu \gg 1$ as well as $P/M \ll 1$. Our analysis clarifies that the condition $\nu \ll 1$ can be relaxed to $\nu > 0$ without changing the general scaling of entanglement-assisted capacity concerning the number of modes $M$ when optical amplifiers are used. In our model, we assume an amplifier which \textit{completely} regenerates the signal while also introducing some noise. This is in contrast to the previous work \cite{jarzyna2019ultimate} and more in line with the analysis of the quantum limit in \cite{ntzel2022operating}, where it has been shown that under the rule of constant amplification (in the sense of amplification is independent of the signal energy), both Shannon and Holevo capacity approach a limiting value when $M\to \infty$. Our results show that the entanglement-assisted capacity can still grow limitless under the same amplification rule when $M\to \infty$.

To interpret our results, it is essential to note that the fundamental assumption of the analysis is that future fiber networks will obey a per-fiber power limit $P$. At the same time, we assume that communication systems will enable increasing orthogonal modes so that the assumption $M\to\infty$ is justified.

\section{System Model}
Following the exposition \cite{banaszek2020quantum}, we use a modeling approach which starts with a narrow band, linearly polarized optical signal coming in the form of uniformly spaced pulses in temporal slots of duration $B^{-1}$. 
Such a system occupies a spectral bandwidth proportional to the slot rate $B$. It can quantum-mechanically be modeled by assigning a quantum state $|\alpha\rangle \in \mathcal F$ to every slot; $\mathcal F$ being the Fock space which has a countably infinite orthonormal basis $\{|n\rangle\}_0^\infty$ called the number-state basis. A system in state $|n\rangle$ consists of exactly $n$ photons. 
A coherent state $|\alpha\rangle$ can be written as
\begin{equation}
    |\alpha \rangle = e^{-|\alpha|^2/2}\textstyle\sum_{n=0}^{\infty}\frac{\alpha^n}{\sqrt{n!}}|n\rangle.
\end{equation}
The expected energy of a coherent state $|\alpha \rangle$ equals $|\alpha|^2$. If these pulses are detected and then converted to electronic values using homodyne detection in a slot-by-slot fashion, and if thermal noise with an average number of $\nu$ photons per slot is present at the receiver, one can model the capacity of a such channel using the Shannon formula:
\begin{equation}\label{eq:C_S}
    C_S(M,P,\tau,\nu) = M \cdot \log \left(1+\tfrac{\tau \cdot P}{M(1+\nu)}\right)
\end{equation}

Motivated by the work \cite{Holevo1973, Holevo1998c}, researchers have also studied the capacity of such an optical data transmission system when arbitrary quantum measurements are allowed at the receiver. When it comes to the concrete realization of this concept, one speaks of so-called \ac{JDR}, where a first proposal was made in \cite{guha11} and a second in \cite{BanaszekPatent}. Both proposals use an interferometric setup to interfere with a series of consecutive pulses (slots) with each other prior to detection. This setup is ideal for harnessing the advantage of \ac{JDR}s. Which is in the region where the received signal energy per slot is very weak. Due to the relation
$P = \mathbb{E}(|\alpha_j|^2)B h f_c$,
it is possible to reach the limit of small received energy per slot by keeping $P$ constant while increasing $B$; where: $P$ is the average optical power of the signal (in watts), $h=\SI{6.626e-34}{\joule\second}$ is Planck constant, $f_c$ is the carrier frequency and the complex numbers $\alpha_j$ specify the symbols that are generated at the sender. The capacity of such channel employing a \ac{JDR} is given by the Holevo capacity:
\begin{equation}\label{eq:C_J}
C_J(M,P,\tau,\nu)=M\cdot \big(g(\tfrac{\tau P}{M}+\nu)- g(\nu) \big),
\end{equation}
where $g(x)=(x+1)\log_2(x+1)-x\log(x)$ is the von Neumann entropy of zero-mean single-mode thermal state of mean photon number $x$. Note that the assumed noise level $\nu$ in equations \eqref{eq:C_J} and \eqref{eq:C_S} is \textit{independent} of the power $P/M$ of the individual pulses. 
Finally, we introduce the third central data transmission concept — entanglement-assisted communication — emerging from the work \cite{bennett1999entanglement}. Entanglement-assisted communication assumes that classical information is not encoded in a quantum state, such as coherent state $|\alpha\rangle$, but rather an arbitrary state shared between the sender and receiver. Therefore, distributing an entangled state must be considered in the system design. The work \cite{noetzel2020gewi} described such a data transmission system as one where idle periods are used to distribute the entangled states, which are then consumed when data needs to be transmitted. The states that can be utilized for this task can, for example, be so-called squeezed states:

\begin{equation}
    |s_r\rangle = \tfrac{1}{\cosh (r)} \textstyle\sum_{n=0}^\infty \tanh^n (r) |n\rangle_I | n\rangle_D.
\end{equation}
In this case, the part of the state indexed with $I$ is transmitted to the receiver at the time 'idle' time. The part indexed with $D$ is used later to encode the data. The capacity of such a channel when the $I$ part of the state is already distributed is given by:
\begin{equation}\label{eq:C_E}
    C_E(M,P,\tau,\nu) = M \sum_{x=0}^1 g\big(\tfrac{\tau^x P}{M} + x\nu\big)-
    g\big(d_x(\tau, \tfrac{P}{M}, \nu)\big).
\end{equation}
Here the functions $d_x$, derived from \cite{guha2020infinite}, are defined via:
\begin{align}
    d_x(\tau, n,\nu) &= \big(d(\tau,n,\nu)-1+(-1)^x((\tau -1)n+\nu)\big)/2\\
    d(\tau, n,\nu) &= \sqrt{((1+\tau)n + \nu + 1)^2-4\tau n(n+1)}
\end{align}

\subsection{Fiber link with optical amplifiers}

\begin{figure}[H]
    \centering
    \includegraphics[]{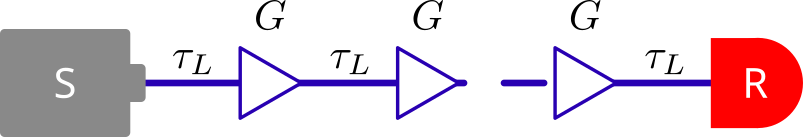}
    \includegraphics[]{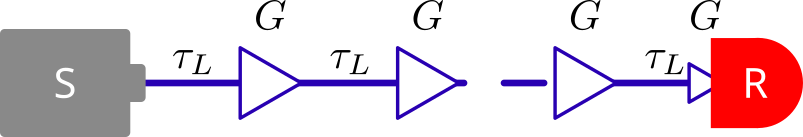}
    \caption{Scheme, representing an amplified segmented link without pre-amplification prior to detection (above) and with pre-amplification prior to detection (below).}
    \label{fig:fiber_links}
\end{figure}

With physical models and capacities defined, we can now illuminate our amplified fiber link. The total length of the fiber link is divided into $K$ segments of equal length $L$. At the end of each segment, the signal is amplified by an optical amplifier, as presented in Figure \ref{fig:fiber_links}. At this level of analysis, the amplifier is completely modeled by its amplification factor $G$. The transmittivity of one segment can be computed by $\tau_L=e^{-\alpha\cdot L}$.
In our case we study the multimode fiber with power loss of \SI{-0.22}{\deci\bel} \cite{donlagic2009low} which using the formula from \cite{czerwinski2022efficiency}, translates to $\alpha\approx0.05$. In our work, we assume common loss for all modes. As authors in \cite{donlagic2009low} noted, the power loss is fiber diameter, mode number and wavelength dependent. In our calculations we simplify the loss, to be the same for all modes and diameters.

As seen in Figure \ref{fig:fiber_links}, we are studying two different implementations. In the first implementation, the attenuated signal is decoded using Passive Receiver (without pre-amplification), while in the second signal is decoded by Active Receiver (with pre-amplification). For each case, the three corresponding capacities for the three receiver types ($X\in\{S, J, E\}$, representing Shannon, Holevo, and entanglement-assisted capacity, respectively) can then be computed:
\begin{equation}
    C_X = C_X(M,P,\tau_{ef}, \nu_{ef}).
\end{equation}

Expressions for effective transmissivity depend on the amplification operation and whether or not an active or passive receiver is used. This study focuses on two amplification models differing in the extent of amplification, namely full signal power regeneration (${G_1= 1/\tau_L}$) and amplification relative to the number of photons per mode transmitted (${G_2=\tfrac{1+n}{1+\tau_L n}}$), as documented in \cite{ntzel2022operating}. Given the amplification models, we write the expression for $\tau_{ef}$. For the passive receiver:
${\tau^{p}_{ef} = \tau_L \textstyle\prod_{i=1}^{K}G\tau_L}$,
and active receiver,
${\tau^{a}_{ef} = \textstyle\prod_{i=1}^{K+1}G\tau_L}$,
where the $G\in\{G_1, G_2\}$ is used depending on the scenario. For the case of full signal regeneration amplification, the $\tau_{eff}$ reduces to $\tau_L$ or $1$ for the passive and active receiver, respectively.

The amplifier injects $G-1$ noise photons into each mode when the signal is amplified. The noisy photons experience attenuation just like the signal photons. With this, the number of noisy photons $\nu_{eff}$ arriving into the receiver can be computed using the following sum of geometric series:
\begin{align}
   \nu_{eff}
        &= \left(\tfrac{\tau_L-\tau_L^{K}}{1-\tau_L}\right)(G-1).
\end{align}
With the use of signal regeneration amplification $G=G_1$ we then expect to receive $\nu_{ef} = 1 - \tau_L^{K-1}$ in the case of passive receiver and $\nu_{ef}=\tfrac{1-\tau_L^K}{\tau_L}$ in the case of active receiver. In the second scenario (using $G_2$ amplification rule) we compute $G-1=(1-\tau_L)\cdot n/(1+\tau_L\cdot n)$. With that, we can finally evaluate the following:
\begin{equation}
    \nu_{ef}(n)= \tfrac{(\tau_L-\tau_L^{K})n}{1+\tau_L\cdot n}, \label{eq:nu_eff_p}
\end{equation}
for passive receivers and
\begin{align}
     \nu_{ef}(n) 
                 &= \tfrac{(1-\tau_L^{K})n}{1-\tau_L \cdot n},\label{eq:nu_eff_a}
\end{align}
for active receivers. As one can see, we have, in both cases, an amount of noise that is strictly larger than zero ($\nu_{ef} > 0$). We also note that the noise experienced by the receiver heavily depends on the choice of amplification, which can easily be deducted from the equations \eqref{eq:nu_eff_p} and  \eqref{eq:nu_eff_a}.
\section{Results and Discussion}
This section discusses the practical feasibility of both \ac{QIP} techniques. We aim to clarify the applicable causality rules guiding the three different capacity scaling laws with amplification models and with active and passive receivers.
More importantly, we illustrate which combination of discussed technologies (amplification and receiver designs) calls for a specific use case. 
Section \ref{sec:power_comparison} will demonstrate that, in general, the $G_1$ amplification rule yields better communication rates than $G_2$ but requires more power for amplification. This limits its use if the unbounded increase in the number of modes is assumed.

In our case, the analysis of functions with multiple variables: the number of segments $K$, length of segments $L$, number of spatial modes $N$, and permutations of the described setups, is complex. Therefore, we have highlighted the most illustrative cases for our discussion.

\subsection{Amplifying with $G_1=1/\tau_L$}
First, we assess capacity scaling for links using the $G_1$ amplification rule. We discuss the scaling in the design with passive and active receivers.

\subsubsection{Decoding with Passive Receiver}
\begin{figure}[H]
    \centering
    \includegraphics[width=0.85\linewidth]{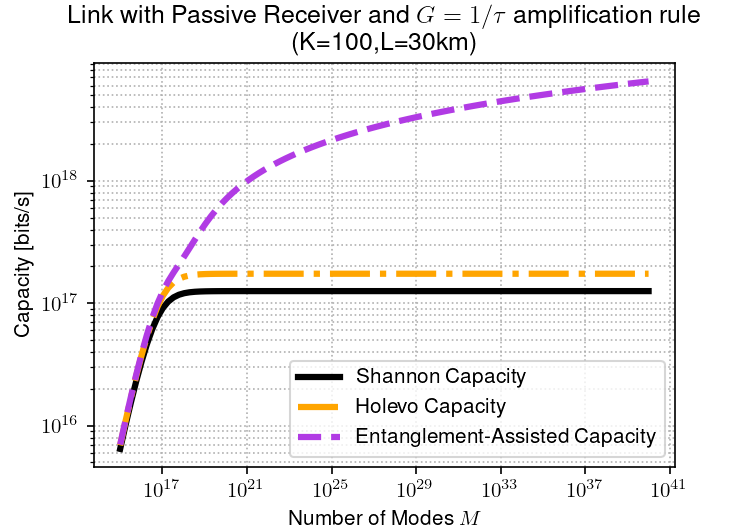}
    \caption{Capacity scaling of the link using passive receiver and $G_1=1/\tau_L$ amplification rule.}
    \label{plt:passive_g1}
\end{figure}

Decoding with a passive receiver nicely outlines the advantage of \ac{QIP} techniques, as seen in Figure \ref{plt:passive_g1}. More importantly it outlines the unbounded scaling of the entanglement-assisted receiver design.

\begin{figure}[h]
    \centering
    \includegraphics[width=0.85\linewidth]{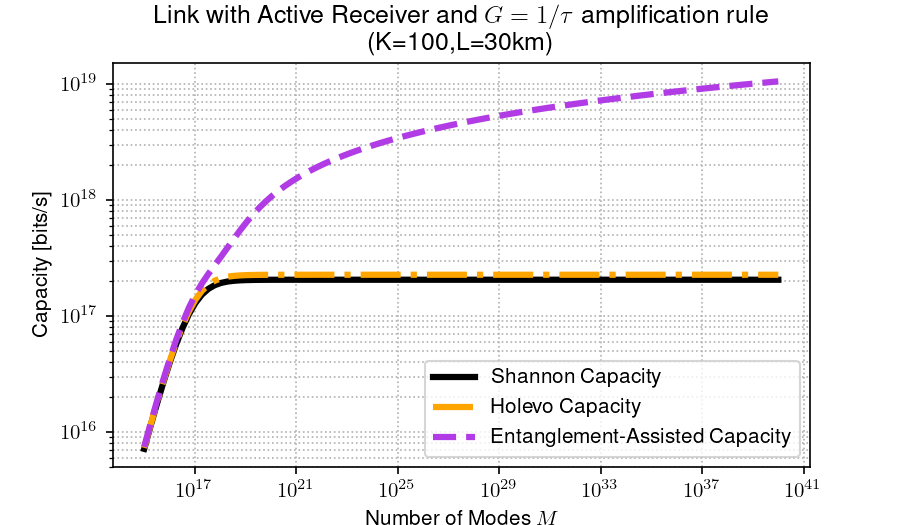}
    \caption{capacity scaling of the link using active receiver and $g=1/\tau_l$ amplification rule. this allows only \textit{entanglement-assisted} capacity to scale logarithmically as $m\to\infty$.}
    \label{plt:active_g1}
\end{figure}

This scenario closely resembles the scenario studied in \cite{ntzel2022operating}. The authors derived a saturation limit for classical communication under similar circumstances in that work. For the scenarios depicted in Figures \ref{plt:passive_g1} and \ref{plt:active_g1}, the bound is computed to be $C_{S}^{(\text{max})}=\tfrac{P \cdot \tau}{\ln 2} \approx \num{3.2e17}$ and as one can see, the bound holds for our case also. With more amplifiers or higher amplification gains $G_2$ resulting from longer fiber segments ($\tau_L$), increasingly more noise is added. This results in the saturation of the Holevo capacities. 

\subsubsection{Decoding with Active Receiver}

Replacing the passive receiver from Figure \ref{plt:passive_g1} with active one improves all capacity rates. Entanglement-assisted capacity is improved the most, seeing almost two-fold gain for the case of $M=\num{1e40}$ as opposed to almost negligible gain seen by other two capacities.

\subsubsection{Derivation of Approximating Formula for $C_E$} 
Our goal here is to derive an approximation to $C_E$, which allows us to study, with high accuracy, its asymptotic scale when the number $M$ of orthogonal channels, which are subject to a global power constraint $\textstyle{\sum_{m=1}^MP_m\leq P}$,
increases. In this case, the scaling of the overall system capacity with $M$ is described by the quantities $C_{X}(M, P,\tau,\nu)$. For $X=C$ and $X=J$ the limit of these quantities has been stated in \cite{ntzel2022operating}, where the formula
$\lim_{M\to\infty}C_{C}(M, P,\tau,\nu)=P\cdot\tau\cdot\log(1+\nu)\nu^{-1}$ can be found with $\tau$ being the transmittivity of the last segment (which is equal to $1$ if active receiver is used) and $\nu$ being equal to $1-\tau^{K-1}$ for passive receiver and $\nu_K=\tau^{-1}-\tau^{K-1}$ for active receiver. This value sets the true benchmark for entanglement-assisted communication to provide an advantage in data transmission capacity. Thus, The question is which values of $(M, P,\tau,\nu)$ the entanglement-assisted capacity can outperform the non-assisted ones by a given factor. 

As one can see from \eqref{eq:C_E}, it is helpful to understand the individual scaling of the involved four terms. In this analysis, it is helpful to study the terms $(g(P/M)- g(d_+(\tau, P/M, \nu)))M$ and $(g(\tau P/M+\nu)- g(d_-(\tau, P/M, \nu)))M$ separately.

Our approximation rests on the following observations: For every $\tau\in[0,1]$ and $\nu\geq0$ it holds $d(\tau,0,\nu)=\nu+1$, which then implies $d_-(\tau,0,\nu) = 0$.

Obviously, it holds $P/M\to0$ as $M\to\infty$. 
Our second observation is that the derivative of $d_-(1,n,\nu)$ at $n=0$ satisfies
\begin{align}
    \partial_n d_-(\tau,n,\nu)|_{n=0}=(1+\nu-\tau)\cdot(1+\nu)^{-1}.
\end{align}
Since $\lim_{\epsilon\to0} g(\nu+\epsilon)=g(\nu)$, $\lim_{M\to\infty} d_+(\tau, P/M,\nu)=\nu$ and $\partial_n d_+(\tau, P/M,\nu)|_{n=0}=\tau\nu/(1+\nu)$ we conclude by L'Hospital's rule that the term $(g(\tau P/M+\nu)-g(d_{+}(\tau, P/M, \nu))M$ asymptotically approaches a constant given by $(1-\tau\nu /(1+\nu))\log(1+\nu)/\nu$. We neglect this constant since we aim to study the asymptotic growth of $C_E$. Thus, the asymptotic growth of $C_E$ must be well approximated by
\begin{align}
    C_E(M,\tau,P,\nu) \approx M\left(g(\tfrac{P}{M})-g(d_-(\tau, \tfrac{P}{M},\nu))\right).
\end{align}
For small values of $\epsilon>0$ it holds $g(\epsilon)\approx(1+\epsilon)\epsilon-\epsilon\log(\epsilon)$. Therefore, setting $T:=(1+\nu-\tau)/(1+\nu)$ and abbreviating $PM^{-1}$ as $\epsilon$ we arrive at

\begin{equation}
\begin{aligned}
   C_{E}&(M,\tau,P,\nu) 
    \approx P \left(g(\epsilon)-g(T\epsilon)\right)/\epsilon\\
    &\approx \tfrac{P}{\epsilon} \left((1+\epsilon)\epsilon-(1+T\epsilon)T\epsilon-\epsilon\log\epsilon+T\epsilon\log T\epsilon\right)\\
    &\approx P \left(-\epsilon\log\epsilon+T\epsilon\log T\epsilon\right)/\epsilon\\
    &\approx P(1-T)\log T\epsilon\\
    &\approx P(1-T)\log M.
\end{aligned}
\end{equation}
Here we subsequently neglected all terms converging to constants. 

This formula applies to both the active and passive receiver, where the difference between the two cases is handled by $1-T$. It can be set in comparison to existing ones, and it shows that entanglement-assisted communication can enable infinite growth even for amplified optical networks. However, when numbers for existing networks are inserted, this formula shows that the parameter regions where entanglement-assisted communication might be helpful have yet to be reached.

\subsection{Amplifying with $G_2=(1+n)/(1+\tau_L n)$ }
Amplification sensitive to the input photons per mode scales also logarithmically but behaves differently due to the severely restricted amount of noise injected by the amplification compared to the previous case $G_1$. The noise injected into the communication link, computed using the expression \eqref{eq:nu_eff_a}, is approaching zero as $M\to\infty$. As can be understood from the gain formula $G_2$, also the gain factor is approaching 1 as $M\to\infty$. Due to both stated factors this amplification rule makes little sense in overall long fiber links, so in this case we showcase shorter links.

\begin{figure}[H]
    \centering
    \includegraphics[width=0.85\linewidth]{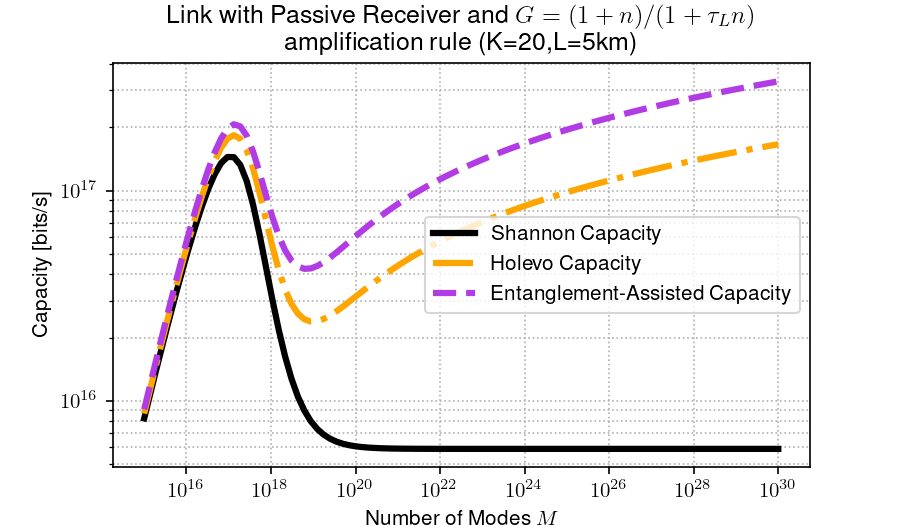}
    \caption{Capacity scaling of the link using a passive receiver and $G_2$ amplification rule. \textit{Shannon} capacity does not benefit in the presence of more modes (after some global maximum has been reached). As $M \to \infty$, both \ac{QIP} capacities scale logarithmically.}
    \label{plt:passive_g2}
\end{figure}

The scaling presented in Figure \ref{plt:passive_g2} depicts how the capacity of the link scales with new modes. In this scenario the Shannon capacity experiences a global maximum, while Holevo and entanglement-assisted capacity experience a local maxima at the same point. This local and global maximum for classical and \ac{QIP} techniques respectively could offer a way of rate cost optimization. The reason for that is that when the number of modes is relatively low, the gain factor stays relatively high, which helps with improving the capacities. With $M\to\infty$ the gain $G_2$ approaches 1, so in limit with this amplification model, we are modeling optical fiber without any amplification. 


\subsection{Comparing power usage between the amplification rules} \label{sec:power_comparison}
\begin{figure}[H]
    \centering
    \includegraphics[width=\linewidth]{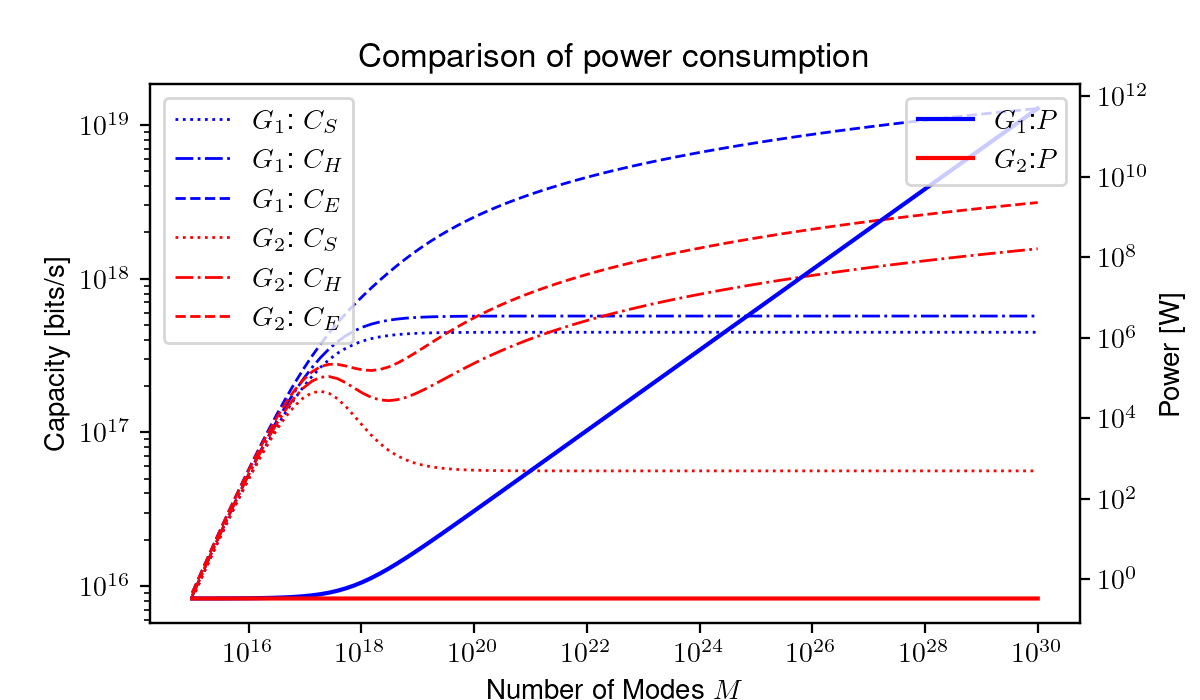}
    \caption{Capacity and power scaling comparison of different amplification techniques. The physical link parameters used for this plot are $L=\SI{10}{km}$ $K=5$. Power scaling should also be considered when considering amplification in quantum-enabled networks.}
    \label{plt:PowerConsumption}
\end{figure}

How do the two different amplification rules affect the power usage of the communication link? We can compute this power consumption by summing all actions of the active components together. The first component is power used by the sender, and then we sum up actions of all of the $K$ or $K+1$ amplifiers, which contribute to the signal ${(G-1)\cdot \tau\cdot n}$ and to the noise part of the pulse $(G-1)$:
\begin{equation}
    \mathcal{P} = n + (K+\tfrac{1}{2} \pm \tfrac{1}{2})\big( (G-1)\cdot \tau n + G - 1   \big).
\end{equation}
Notice that the units are in photons per mode per second, whereas in Figure \ref{plt:PowerConsumption}, the units are converted to Watts by multiplication by the number of modes $M$ and energy of a photon $E=hf_c$.

Obviously, in our amplification models, the use of either model depends on other factors, such as segment length and number of segments as well as signal input power. In the $(G=1/\tau)$ amplification model, much effort is put into restoring the signal, producing much noise. This further results in the power consumption that depends on the number of modes. With this approach, one could quickly achieve power in the system above the given limits, as seen in Figure \ref{plt:PowerConsumption}. 

The scenario is entirely different in the second case, with amplification modeled as $G=\tfrac{1+n}{1+\tau n}$. We amplify by a lower mode-dependent gain factor, which scales down with additional modes. While per mode capacity will always be lower in this scenario, we are not inflating our power consumption with additional modes. This amplification model also operates below the fiber power limits for any number of modes.


\subsection{Discussion}
The analysis outlined the operational benefits and drawbacks of using any one amplification rule in different configurations. While the power scaling of the first amplification rule might not be favorable, future multimode fibers are yet to be developed, resulting in possible higher power limits. With such advances, parameter regime overlap of first amplification rule $G_1$ might offer a viable path to \ac{QIP} achievable capacity gains.
If power remains a limiting factor, then the second amplification rule $G_2$ will offer a better alternative. In this case, however, the existence of maxima (as seen in Figures \ref{plt:passive_g2} and \ref{plt:PowerConsumption}), might offer a cost-optimized capacity gain.
Note that the development of amplifiers for multimode fibers are a field of active research.
\section{Conclusions}
We studied the capacity advantage of entanglement-assisted data transmission in amplified multi-mode fiber links. Our work is the first to provide a technological argument linking this concept to the development of novel fiber technologies which are relevant to public communication networks.

We highlighted the impact of different amplification rules on the scaling of the link capacity when the number of (spatial) modes increases. 
Our analysis of both active and passive receivers revealed that the choice of the receiver does not generally affect capacity scaling.

In addition
, we 
analyzed the power consumption of amplifiers under different amplification rules, recognizing that the amount of power that optical fiber can transport is finite. 
Future work will need to address open questions like the accurate modelling of wave propagation in the fiber, cross-talk or entanglement consumption of the link, and come forward with designs for multi-mode amplifiers. 

\balance

\section*{Acknowledgments}
The authors thank Prof. Dr.-Ing. Norbert Hanik for fruitful and insightful discussion regarding the fiber mode scaling.

Funding from the Federal Ministry of Education and Research of Germany /BMBF) via grants 16KISQ039, 16KIS1598, and ”Souver\"an. Digital. Vernetzt.” joint project "6G-life", grant number: 16KISK002, DFG Emmy-Noether program under grant number NO 1129/2-1 (JN) and the support of the Munich Quantum Valley (MQV) and Center for Quantum Science and Technology (MCQST) are acknowledged.
\printbibliography
\end{document}